\documentclass[preprint,aps,showpacs,floatfix]{revtex4}
\usepackage{graphicx}
\usepackage{graphics}
\usepackage{psfrag}
\usepackage{color}      
\usepackage{subfigure}

\newcommand{\vv}{\mathbf{v}}
\newcommand{\xhat}{\hat{\mathbf x}}
\newcommand{\rr}{\mathbf{r}}
\newcommand{\vvv}{\mathbf{\bar v}}

\begin{document}
\title{Turbulent superfluid profiles in a counterflow channel}
\author{L. Galantucci$^{1,2}$, C.F. Barenghi$^2$, M. Sciacca$^{3,2}$, 
M. Quadrio$^1$ and P. Luchini$^4$}
\affiliation{
$^1$Dipartimento di Ingegneria Aerospaziale, Politecnico di Milano,
          via La Masa, 34 20156 Milano, Italy,\\
$^2$School of Mathematics and Statistics, 
Newcastle University, Newcastle upon Tyne, NE1 7RU, England, UK,\\
$^3$Dipartimento di Metodi e Modelli Matematici, Universit\`a di Palermo,
          Viale delle Scienze, I-90128 Palermo, Italy,\\
$^4$Dipartimento di Ingegneria Meccanica, Universit\`a di Salerno,
          via Ponte don Melillo, 84084 Fisciano (Salerno), Italy,
}

\date{\today}
\begin{abstract}
We have developed a two-dimensional model
of quantised vortices in helium~II moving under the influence of applied normal
fluid and superfluid in a counterflow channel. We predict
superfluid and vortex-line density profiles which could be experimentally
tested using recently developed visualization techniques.
\end{abstract}
\maketitle

\section{Motivation}
\label{sec:1}

The recent development of visualization techniques in superfluid
helium based on micron-size tracers \cite{Bewley,Zhang-VanSciver} 
has raised the possibility of experimentally determining
superfluid and normal fluid profiles and the spatial distribution of 
quantised vortices in a channel, thus solving outstanding problems
in quantum turbulence. Particularly interesting (and relevant to
engineering applications) is
the turbulence induced by heat transfer (counterflow
turbulence).  It is well known that if the applied heat
flux $\dot Q$ is less than a small critical value ${\dot Q}_c$ then
the heat
is carried by the normal fluid component, and the superfluid component
flows in the opposite direction to conserve mass. 
If ${\dot Q}>{\dot Q}_c$ the superfluid component becomes turbulent,
forming a disorganized tangle of quantised vortices. In some geometries,
at larger heat flux a transition to a more intense vortex tangle 
has been observed \cite{Tough}, which is perhaps related to the onset of
turbulence in the normal fluid \cite{Melotte}.

In this paper we are concerned with the intermediate heat transfer
regime, in which the normal fluid is still laminar, but the superfluid
forms a turbulent tangle. The issue which we address is the average
superfluid profile and the average spatial distribution of vortices in the channel.
Let $\vv_n$ and $\vv_s$ be respectively the normal fluid and superfluid velocity fields.
The normal fluid satisfies no slip boundary conditions at the wall
of the channels, so, for the sake of simplicity, we assume that $\vv_n$
is a steady classical parabolic Poiseuille profile (this approximation implies
that the constant pressure drop along the channel is the same with and without
the vortex tangle). The superfluid slips at the boundaries, so, in the 
absence of vortices, it is natural to assume that $\vv_s$ has a constant
(uniform) profile. The velocity fields
$\vv_n$ and $\vv_s$ are related by the counterflow condition 
of no net mass flow along the channel.

The question which we ask is the following: 
if ${\dot Q} > {\dot Q}_c$, are the vortices uniformly distributed
in the channel, or do they organize themselves spatially, creating a
non-uniform superfluid profile at scales larger than the average
vortex separation $\ell$ but smaller than the channel size $D$ ?

\section{Model}
\label{sec:2}

To answer the question we consider the following idealised two-dimensional
model which, we argue, captures the most important physical
ingredients. Let $x$ and $y$ be respectively directions along
and across the channel, with walls at $y=\pm D/2$ and periodic
boundary conditions at $x=0$ and $x=\lambda$. The normal fluid
velocity is $\vv_n=(v_{n}^x,v_{n}^y)=\left ( -V_{n0} [1-(2y/D)^2]\, , \, 0 \right )$ 
with $V_{n0}>0$ and thus pointing in the negative $x$ direction.
The superfluid velocity $\vv_s=(v_{s}^x,v_{s}^y)$ 
is decomposed
in two parts, $\vv_s=\vv_{s0} + \vv_{si}$; the former is the uniform
flow $\vv_{s0}=V_{s0}(t)\xhat$ , $\xhat$ being the unit vector along 
$x$, and the latter,
$\vv_{si}$, is the velocity field induced by 
$N$ vortex points located at positions $\rr_j=(x_j(t),y_j(t))$ 
(for $j=1,\cdots N$) where $t$ is time. Half the vortices have
positive circulation $\Gamma_j=\kappa$, and half have negative
circulation $\Gamma_j=-\kappa$, where $\kappa=10^{-3}~\rm cm^2/s$
is the quantum of circulation in superfluid $^4$He.
To satisfy the superfluid's boundary condition that $v_{s}^y=0$ at
$y=\pm D/2$ (no flow into the wall),
we attach to each vortex point an infinite series of image vortex 
points in the positive and negative regions $y>D/2$ and 
$y<-D/2$ \cite{Greengard}.

The equation of motion of a vortex located at $\rr_j$ is \cite{Schwarz}
\begin{equation}
\frac{d\rr_j}{dt}=\vv_s(\rr_j)  
+ \alpha {\bf s}'(\rr_j) \times (\vv_n(\rr_j) - \vv_s(\rr_j)) + 
\alpha' (\vv_n(\rr_j) - \vv_s(\rr_j)),
\label{eq:Schwarz}
\end{equation}
where ${\bf s}'_j$ is the unit vector along the vortex $j$
(in the positive or negative $z$ direction) and $\alpha$ and $\alpha'$ are 
temperature dependent mutual friction coefficients \cite{BDV}.

The quantity $V_{s0}$ is determined at each time $t$
by imposing the counterflow condition

\begin{equation}
\rho_n <v_{n}^x> + \rho_s <v_{s}^x>=0,
\label{eq:counterflow}
\end{equation}

\noindent
where $\rho_n$ and $\rho_s$ are the temperature-dependent normal 
fluid and superfluid densities, $\rho=\rho_n+\rho_s$ is the total 
helium density,
and $<v_{n}^x>$ and $<v_s^x>$ are the channel averages of the 
$x$-components of 
the normal fluid and superfluid velocities, defined as

\begin{equation}
<v_{s}^x>=\frac{1}{\lambda D}\!\!\int_0^\lambda\!\!\int_{-D/2}^{D/2} v_{s}^x(x,y)dxdy
        =V_{s0}+\frac{1}{\lambda D}\!\! \int_0^\lambda\!\!\int_{-D/2}^{D/2} 
v_{si}^x(x,y)dxdy
\label{eq:average-vs}
\end{equation}

\noindent
and 
\begin{equation}
<v_{n}^x>=\frac{1}{D}\int_{-D/2}^{D/2} v_{n}^x(y) dy
        =-\frac{2}{3}V_{n0}. 
\label{eq:average-vn}
\end{equation}

\noindent
To model the creation and the destruction of vortices within our 2-dimensional
model, we proceed as follows. When the distance between
two vortex points of opposite circulation becomes smaller than a critical
value $\epsilon_1$, we perform a ``numerical vortex
reconnection'' and remove these vortex points; 
similarly, when the distance between a vortex point and a boundary 
is less than $\epsilon_2=0.5\,\epsilon_1$, we
remove this vortex point (the vortex of opposite circulation being the nearest image 
vortex beyond the wall). To maintain a steady state, when a vortex point
is removed, a new vortex point of the same circulation is re-inserted into
the channel, either on the axis ($y=0$) or near the walls
or randomly, as described in the next section.

At selected times, given the vortex configuration $\rr_j(t)$ ($j=1,\cdots N$),
we define a {\it coarse-grained} superfluid velocity $\vvv_s$
by averaging the components of the (microscopic) velocity $\vv_s$ 
over channel strips of size $\Delta$ in the $y$ direction, such that 
$\ell < \Delta <D$. 
The limit $\ell << \Delta << D$ corresponds to the 
Hall - Vinen - Bekarevich - Khalatnikov (HVBK) equations 
\cite{Hills-Roberts}. 
We assume that the velocity $\vvv_s$ is parallel to the walls, 
$\vvv_s=(\bar{v}_s,0)$.  The curl of 
$\vvv_s$ can be interpreted as the
coarse-grained superfluid vorticity.

To make connection with the experiments, we interpret $n=N/(\lambda D)$
(number of vortex points per unit area) as the vortex line density $L$
(vortex length per unit volume), from which $\ell=n^{-1/2}$ is the
average intervortex spacing. The imposed heat flux 
$\dot Q$, which is reported in experiments, is related to
the average counterflow velocity $V_{ns}=<v_{n}^x>-<v_{s}^x>$ by the 
relation ${\dot Q}=T \rho_s S V_{ns}$, 
where $T$ is the temperature and $S$ the specific entropy.

The calculation consists in computing the evolution of the vortex
points, starting from an arbitrary initial condition, until a steady
state regime is achieved and the profile of $\vvv_s$ becomes constant.
The time integration is performed using the second--order Adam-Bashfort 
method with time step $\Delta t$.

\section{Results}
\label{sec:3}

The numerical code solves the governing equations written in
dimensionless form. The units of length, velocity and time
are respectively $\delta_c=D/2=4.55\times 10^{-3}~\rm cm$ , 
$u_c=~\kappa/(2\pi\delta_c)= 3.49\times 10^{-2}~\rm cm/s$, 
$t_c=\delta_c/u_c=0.13~\rm s$. 
Non-dimensional quantities are denoted by the superscript `$\ast$'. 
We choose parameters taking into account the available computing
power and the experiments of Tough and collaborators
\cite{Ladner-Tough,Martin-Tough}: $N=1876$, $\lambda^\ast=6$,
$T=1.7~\rm K$, $D^\ast=2$ (corresponding to tube $R4$ in 
Ref.~\cite{Ladner-Tough}), $\epsilon_1=1.25\times10^{-2}\,\ell$, 
$\Delta=1.25\,\ell$, $V_{ns}^\ast=-478.5$ and 
$\Delta t^*=1.9\times 10^{-6}$. At this temperature, taking into account 
pressure and temperature variations experimentally measured 
along the channels \cite{Ladner-Tough}, we can assume the constant values
$\rho_s=0.112~\rm g~cm^{-3} $, $\rho_n=~3.32\times10^{-2}~\rm g~cm^{-3}$
and $S=~0.395~\rm J~(g~ K)^{-1}~$ \cite{Donnelly-Barenghi}. Using
Eqs.~(\ref{eq:counterflow}), (\ref{eq:average-vn}) and the definition of 
$V_{ns}$ we have $V_{n0}^\ast=553.6$. 
The vortex density is 
$n^\ast=~156.3$, and the average intervortex spacing,
$\ell^\ast \approx 0.08$, corresponds to 
the dimensionless number $L^{1/2}D=D^\ast\sqrt{n^\ast}=~25$ , 
which is typical of counterflow experiments \cite{Martin-Tough}.

\begin{figure}[htbp]
\hspace{-1.5cm}
     \begin{minipage}{0.46\textwidth}
      \centering
       \includegraphics[width=1.0\textwidth]{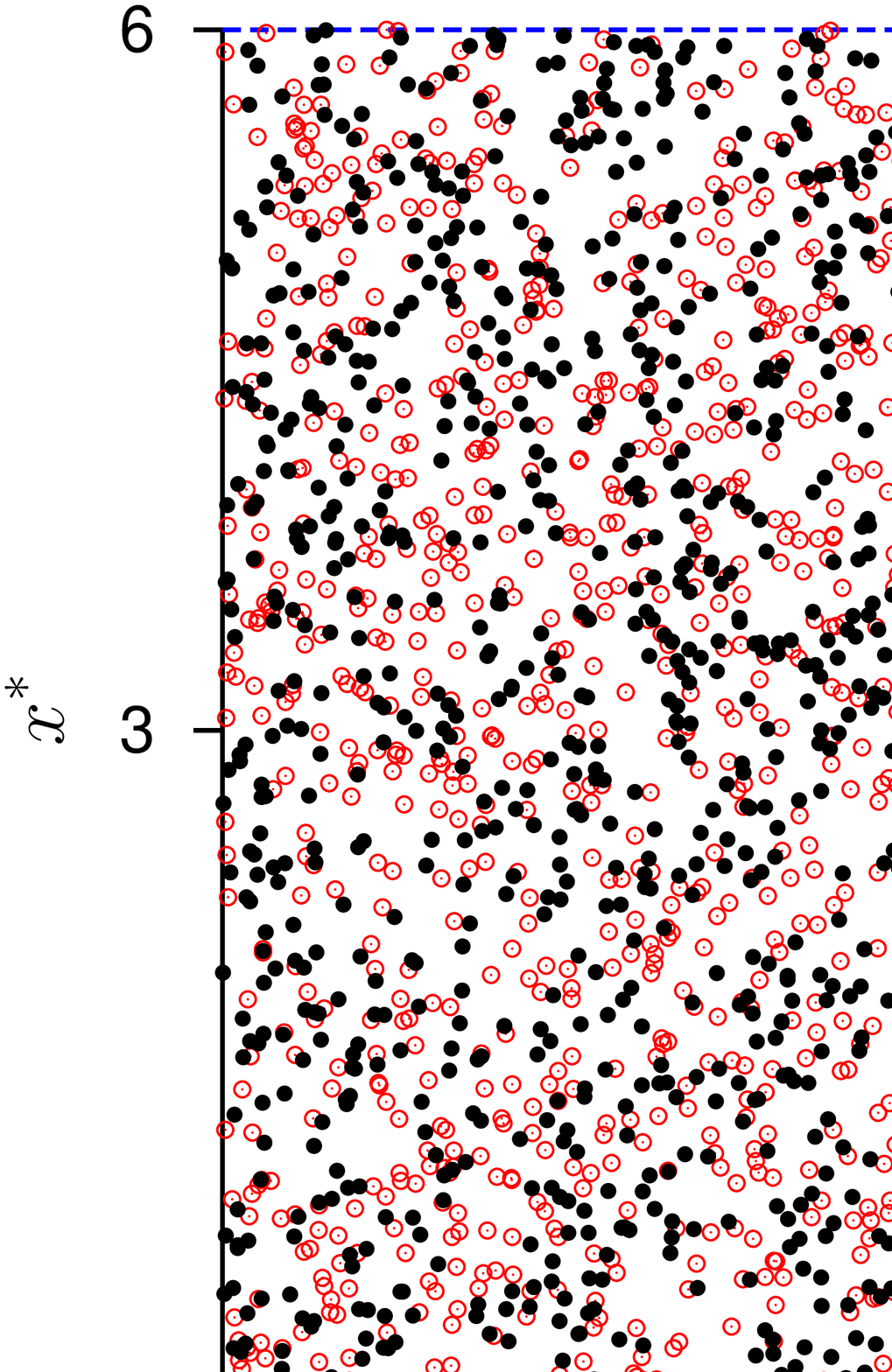}
     \end{minipage}
     \hspace{0.2cm}
     \begin{minipage}{0.46\textwidth}
      \centering
       \includegraphics[width=1.0\textwidth]{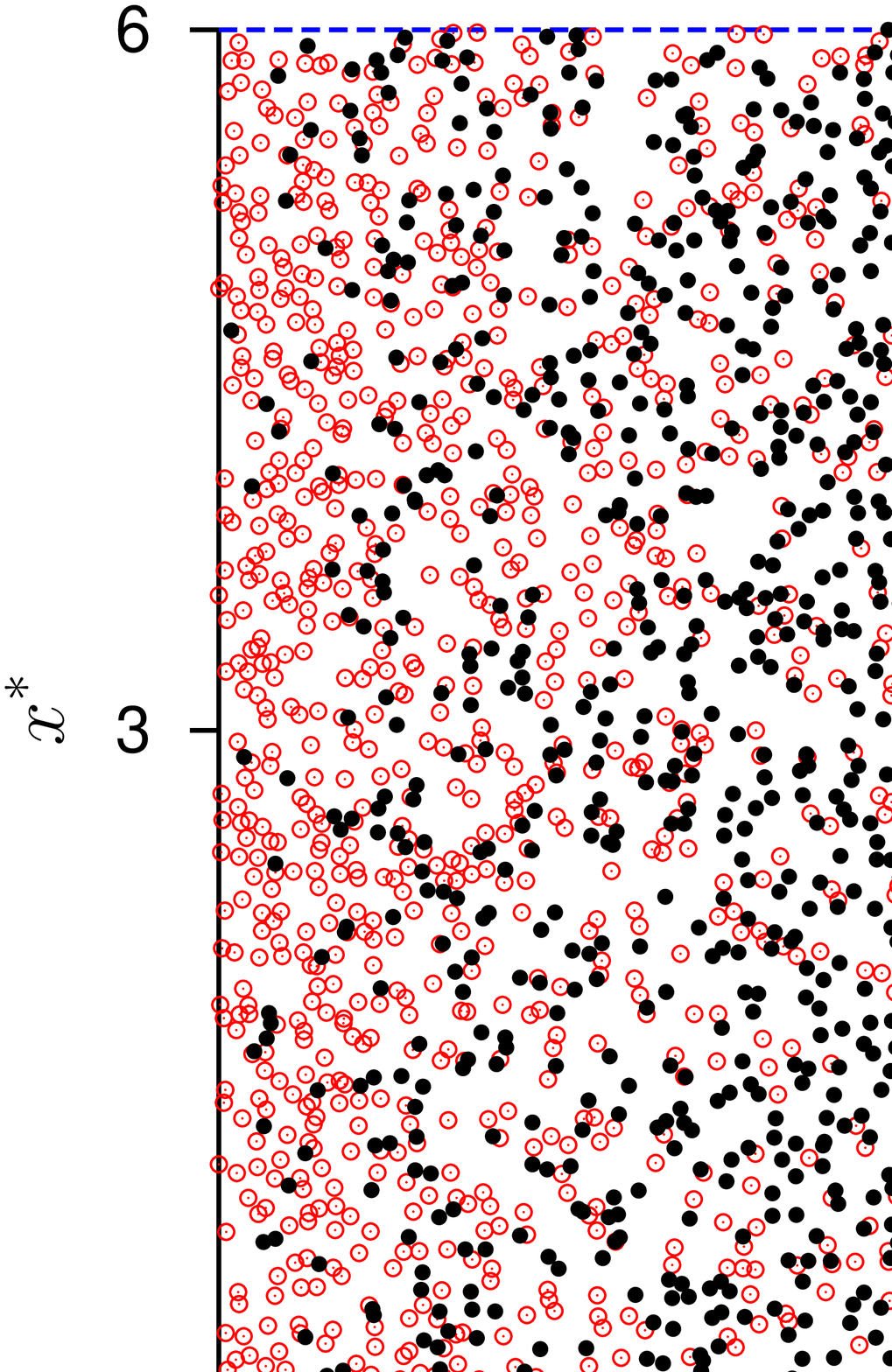}
     \end{minipage}
\vspace{1.1cm}
\caption{(color online). Initial random vortex configuration (left) and 
vortex configuration at $t^\ast=4.75\times 10^{-2}$ (right).
Positive and negative vortices are respectively denoted by empty (red) and 
filled (black) circles.} 
\label{fig:1}
\end{figure}

To distinguish the three different ``re--nucleation'' regimes, 
hereafter we indicate with $(a)$ random vortex-reinsertion in the
channel, and with $(b)$ and $(c)$  vortex re-insertion  
on the axis and near the walls, respectively. 

Case $(a)$ is the simplest and corresponds physically to 
3-dimensional superfluid vorticity production 
taking place throughout the vortex-tangle due to reconnections 
and vortex-rings emission.
The initial condition, shared by all three cases, is shown in 
Fig.~\ref{fig:1} (left) and consists of
a random vortex configuration. During the evolution of case $(a)$, 
positive and negative vortices move on the average towards
the $y^\ast=-1$  and $y^\ast=+1$ wall respectively. The trajectory
of an individual vortex can be very irregular, due to the interaction
with other vortices. A drift in the
positive $x$ direction is superimposed to this motion towards walls.
 
\begin{figure}[htbp]
\hspace{-2.2cm}     
     \begin{minipage}{0.48\textwidth}
      \centering
       \includegraphics[width=1.15\textwidth]{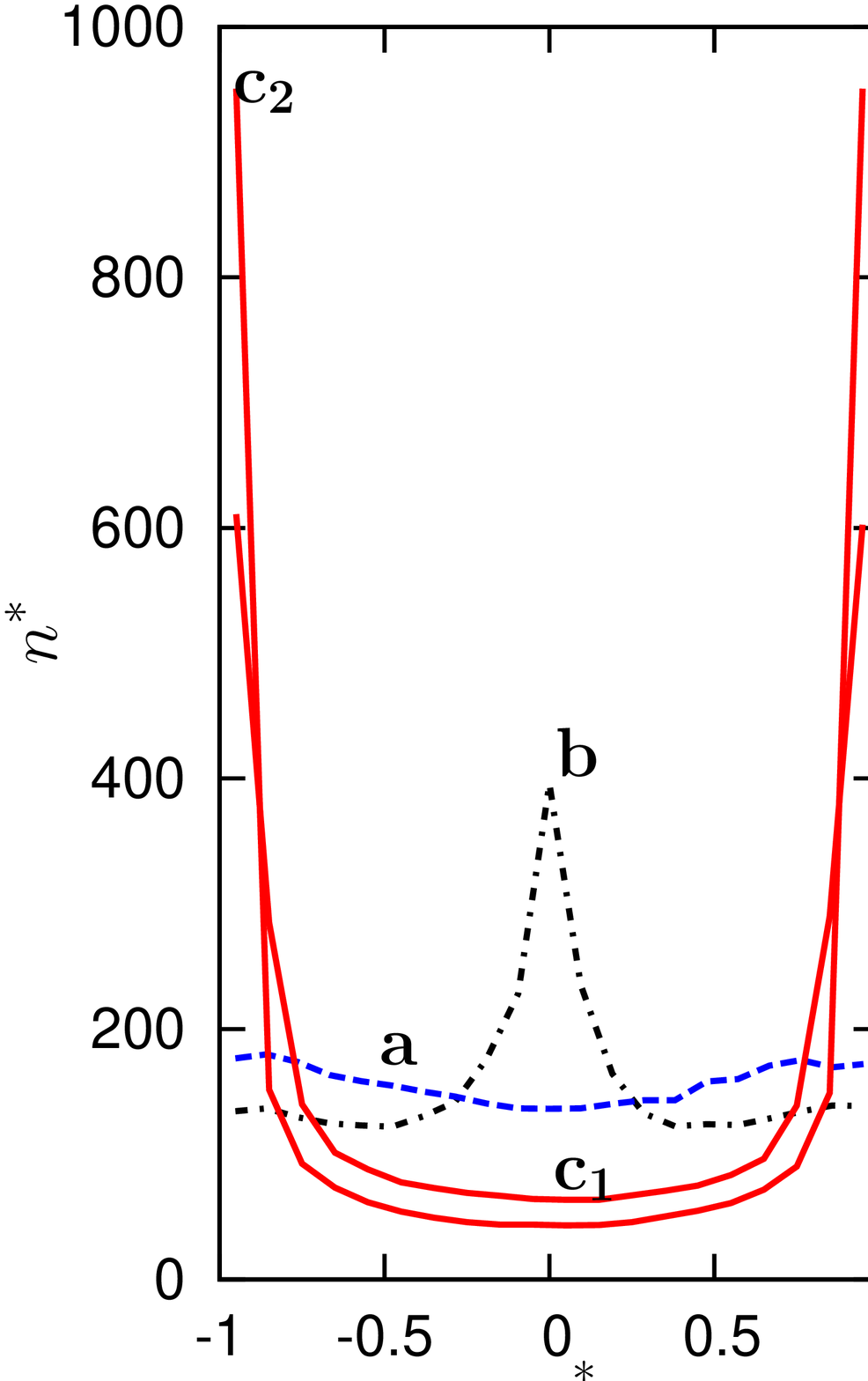}
     \end{minipage}
     \hspace{0.2cm}
     \begin{minipage}{0.48\textwidth}
      \centering
       \includegraphics[width=1.15\textwidth]{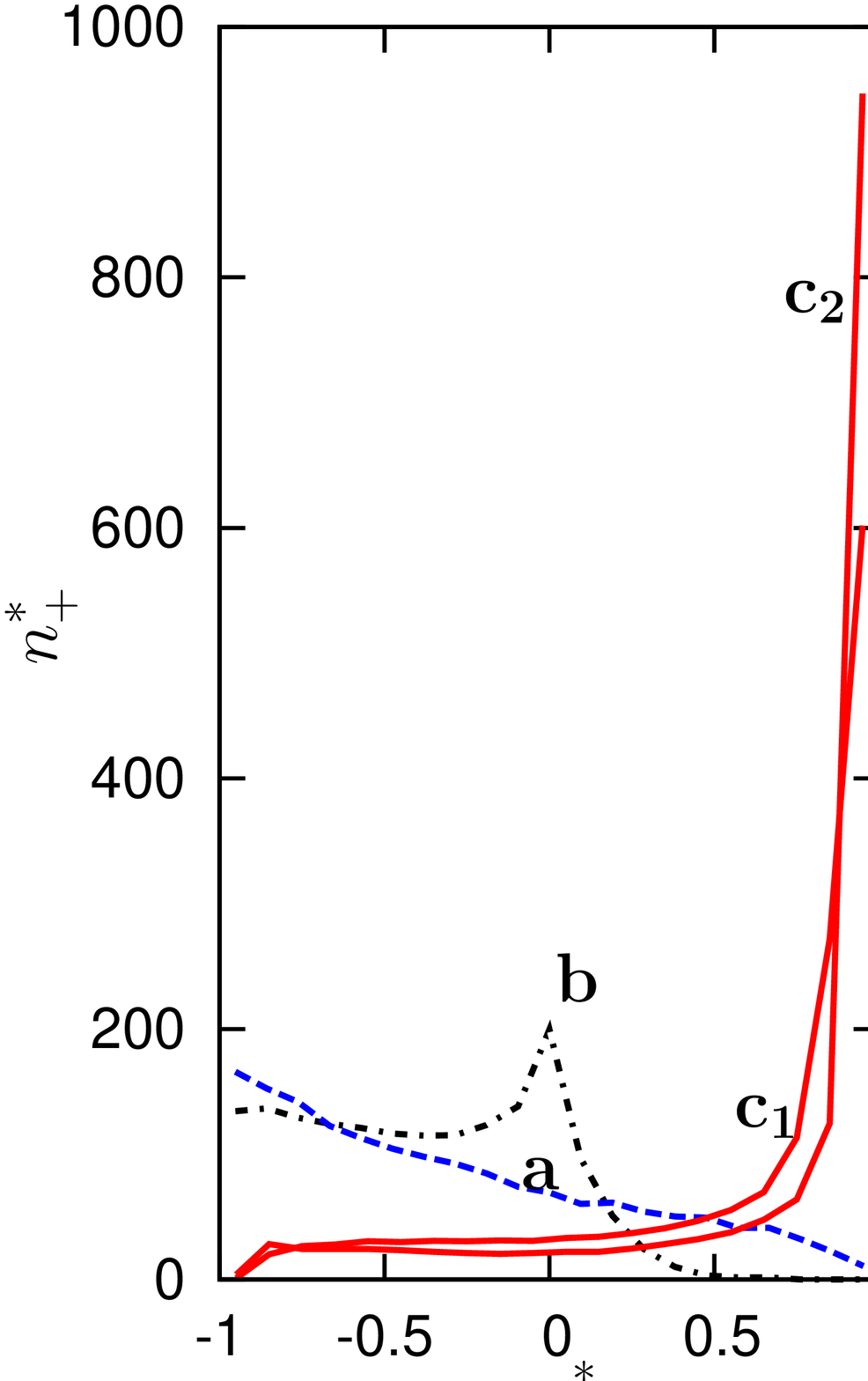}
     \end{minipage}
\vspace{0.2cm}
\caption{(color online). Vortex density $n^\ast(y^\ast)$ (left) 
and positive vortex density $n_+^\ast(y^\ast)$ (right)
for cases: $(a)$, dashed (blue) line; $(b)$, dot-dashed (black) line; 
$(c_1)$ and $(c_2)$, solid (red) lines.}
\label{fig:3}
\end{figure}

After a time-interval of the order of 
$\tau^\ast_{a} \approx 2\times 10^{-2}$, the vortex configuration
reaches a steady state, as illustrated in Fig.~\ref{fig:1} (right). 
Note that there are more positive
vortices in the $y^\ast~<~0$ region and more negative vortices in the  
$y^\ast~>~0$ region, as confirmed
by the density profile of positive vortices, $n^\ast_+(y^\ast)$,
 shown in Fig.~\ref{fig:3} (right) 
(the density profile of negative vortices, $n^\ast_-(y^\ast)$,
is symmetrical with respect to the channel axis). The polarization
of the vortex configuration is not complete - not {\it all} positive
vortices are in $y^\ast~<~0$ and not {\it all} negative vortices
are in $y^\ast~>~0$ - in agreement with arguments discussed in
Ref.~\cite{Hulton}.

\begin{figure}[htbp]
\hspace{-1.9cm}
     \begin{minipage}{0.48\textwidth}
      \centering
       \includegraphics[width=1.1\textwidth]{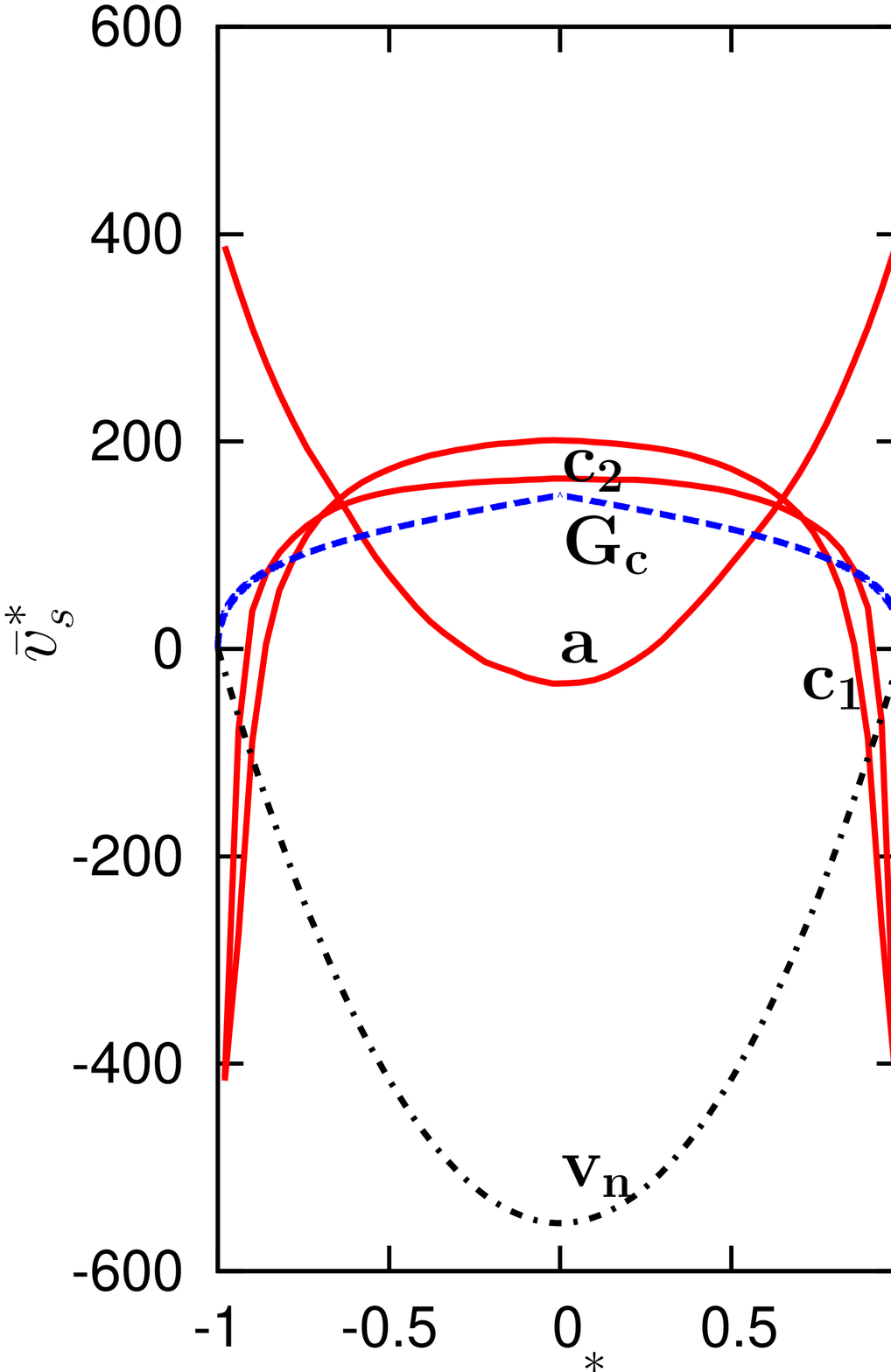}
     \end{minipage}
     \hspace{0.2cm}
     \begin{minipage}{0.48\textwidth}
      \centering
       \includegraphics[width=1.1\textwidth]{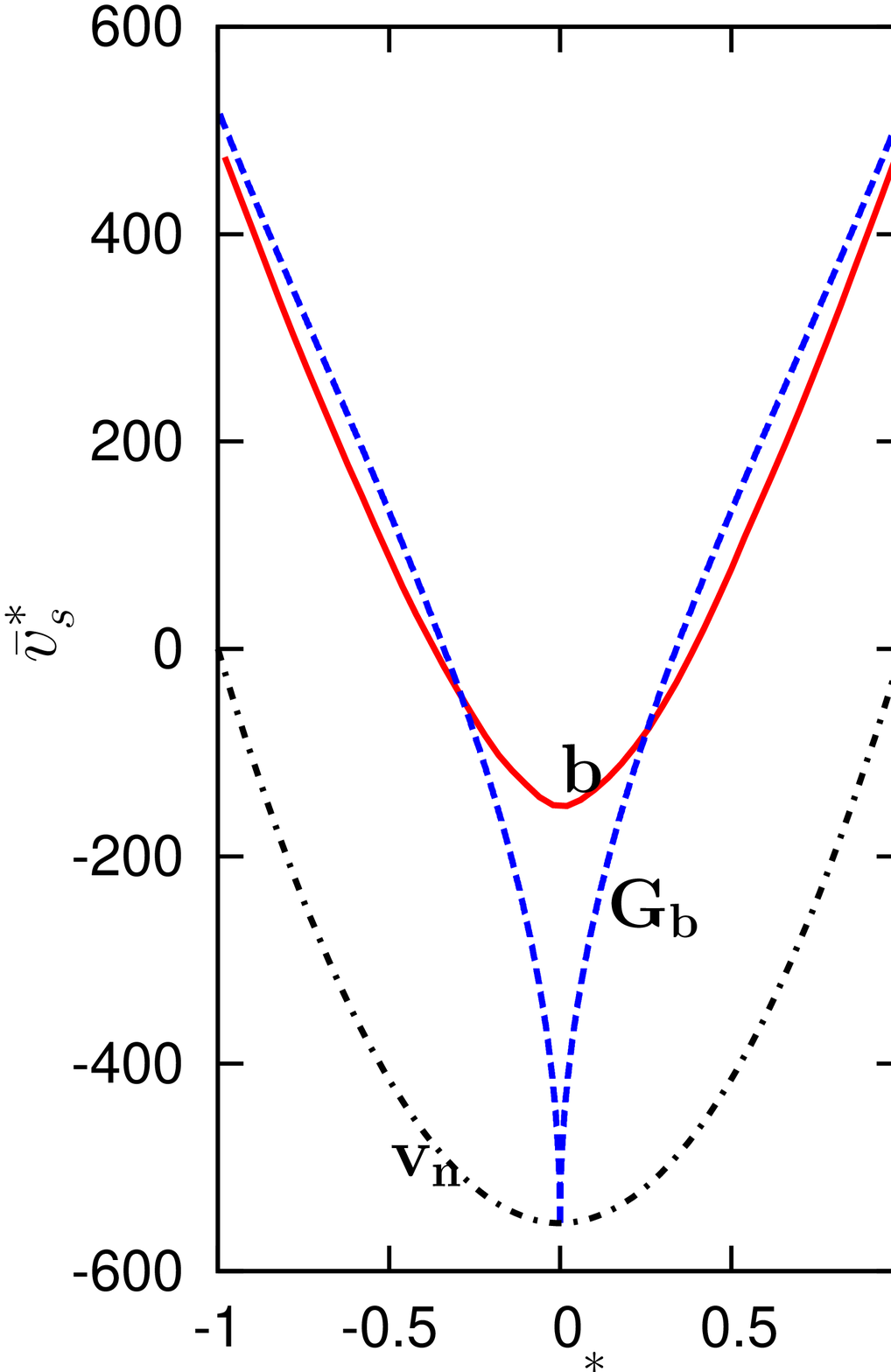}
     \end{minipage}
\vspace{0.2cm}
\caption{(color online).
Solid (red) line: coarse-grained superfluid velocity 
$\bar{v}_s^\ast(y^\ast)$; the
labels $(a)$, $(c_1)$, $(c_2)$ (left) and $(b)$ (right) correspond to 
the cases discussed in the text. Dashed (blue) line: analytical
laminar solution \cite{GCM} of HBVK equations which applies Geurst's
approach \cite{Geurst} to Cartesian geometry. Curves $G_c$ (left) and
$G_b$ (right) correspond respectively to cases $(c)$ and $(b)$ in the text.
Dot-dashed (black) line: normal fluid velocity profile (left, right).}
\label{fig:2}
\end{figure}

The resulting coarse-grained superfluid velocity 
profile associated to this partial polarization is compared to the driving 
normal fluid velocity in Fig.~\ref{fig:2} (left): we find that the profile is
almost parabolic, $\bar{v}_s^\ast\sim~y^{\ast^{2.19}}$. 
The distribution of vortices irrespective of their sign,
$n^\ast(y^\ast)$, is almost constant, as shown in
Fig.~\ref{fig:3} (left).

The  other two cases which we investigate, case $(b)$ (nucleation
at $y^\ast=0$) and case $(c)$ (nucleation at
$y^\ast=\pm 1$), are suggested by the analysis of
Geurst \cite{Geurst} based on the HVBK equations. 
Case $(b)$ induces a vortex density profile 
$n^\ast(y^\ast)$ with a sharp peak near $y^\ast=0$ 
(Fig.~\ref{fig:3} (left) ). We find that this concentration
of vortices on the axis is unstable: collective motion 
of vortices of the same polarity appears, and the resulting
coarse-grained superfluid velocity profile,
$\bar{v}_s^\ast(y^\ast)$, shown in 
Fig.~\ref{fig:2} (right), undergoes 
small-amplitude but persistent oscillations. 
A power-law regression yields the dependence 
$\bar{v}_s^\ast\sim~y^{\ast^{1.6}}$; as in case $(a)$, after
a transient of the order of $\tau^\ast_{b}\sim 6\times 10^{-2}$,
we obtain the partially polarized steady state vortex--distribution 
shown in Fig.~\ref{fig:3} (right).

Finally, we consider case $(c)$.
This nucleation regime corresponds physically 
to 3-dimensional superfluid vorticity production arising from 
vortex--rings emission by vortices pinned to the walls. 
In our model, the positive (negative) vortex points are re-inserted near 
(rather than at) the 
$y^\ast=1$ ($y^\ast=-1$) wall, in order not to collapse rapidly on 
the boundaries due to the Magnus-mutual friction forces balance. 
Let $\xi$ be the distance of nucleation away from the walls.
To investigate the dependence of the flow on $\xi$,
we choose values $\xi_1=\ell$ and $\xi_2= 0.5\, \ell$, 
which we refer to as cases $(c_1)$ and $(c_2)$. 
The resulting coarse-grained superfluid velocity profile 
$\bar{v}_s^\ast(y^\ast)$ in the two cases is shown
in Fig.~\ref{fig:2} (left): note the reversed concavity, compared to
cases $(a)$, $(b)$ and the normal fluid.

Fig.~\ref{fig:3} shows that the nucleation location 
clearly influences the  
vortex distribution and the steady state vortex density profile 
$n^\ast(y^\ast)$, which is reached 
after a transient time interval of the order of
$\tau^\ast_{c}\approx 3\times 10^{-2}$, 
independently of $\xi$. 
  
\vspace{-0.5cm}

\section{Discussion}
\label{sec:4}

The model which we have presented, being 2-dimensional, is
clearly rather idealized, and we do not claim that it is
possible to make direct comparison with experiments.
Nevertheless, the model contains what we argue are the most
important physical ingredients, and it allows us to make
predictions about the vortex distribution and the superfluid
profile across a channel at small values of the applied heat flux.
The 2-dimensional 
solution which we have found can be interpreted in 3-dimensions
as vortex loops which move from the outer parts of the channel towards
the centre, speeding up during this process, or, perhaps more precisely,
as a tangle which is polarised by the presence of such loops.

Our coarsed-grained superfluid velocity 
profile $\bar{v}_s^\ast(y^\ast)$ compares very well with the
2-dimensional laminar solution of the HVBK equations  which
can be derived by
applying Geurst's  approach \cite{Geurst} for a cylindrical pipe
to Cartesian geometry \cite{GCM}. 
Indeed, Fig.~\ref{fig:2} (right) shows that case $(b)$ and
this laminar solution \cite{GCM} are very similar
(with the exception of the
channel axis, where Guerst's solution is singular and
$n^\ast \to \infty$).

In Fig.~\ref{fig:2} (left) we compare the 
analytical solution corresponding to nucleation at the walls 
($\xi=0$) to 
the velocity profile $\bar{v}_s^\ast(y^\ast)$ for case $(c)$. 
It is apparent  that, with decreasing $\xi$, $\bar{v}_s^\ast(y^\ast)$ 
tends to the analytical solution (the difference which is present 
near the walls arises from the boundary conditions  
for the superfluid velocity which ensure
infinite vorticity in the nucleation region \cite{Geurst,GCM}).

Further work will generalise the approach which we have presented,
by solving self-consistently the equation for the normal fluid in the
presence of the mutual friction, rather than assuming a 
given profile for $\vv_n$. We also plan to investigate the 
dependence of the profiles on the
vortex densities.
The current rapid progress of visualization techniques, such as the
recently developed laser-induced fluorescence of metastable 
molecules \cite{Vinen}, will clearly stimulate more work
on the nature of laminar and turbulent profiles of helium~II in
channels.

\begin{acknowledgements}
M.S. acknowledges a travel grant of the Istituto Nazionale di Alta
Matematica. C.F.B.'s work is supported by the Leverhulme Trust.
\end{acknowledgements}

\end{document}